\documentclass[a4paper]{jpconf}

\usepackage{amsmath,amssymb}
\usepackage{cite}

\newcommand{\LT}{\left}
\newcommand{\RT}{\right}
\newcommand{\Dmq}{\Delta m^2}
\newcommand{\Eps}{\varepsilon}
\newcommand{\Ie}{\textit{i.e.}}
\newcommand{\diag}{\mathop\mathrm{diag}\nolimits}

\begin{document}

\title{New interactions: past and future experiments}

\author{Michele Maltoni}

\address{Departamento de F\'isica Te\'orica \& Instituto de F\'isica
  Te\'orica UAM/CSIC, Facultad de Ciencias C-XI, Universidad
  Aut\'onoma de Madrid, Cantoblanco, E-28049 Madrid, Spain}

\ead{michele.maltoni@uam.es}

\begin{abstract}
    In this talk I will review the present status and future
    perspectives of some popular extensions of the conventional
    three-neutrino oscillation scenario, from a purely
    phenomenological point of view. For concreteness I will focus only
    on three specific scenarios: non-standard neutrino interactions
    with matters, models with extra sterile neutrinos, and neutrino
    decay and decoherence.
\end{abstract}

\section{Introduction}

Most of the talks presented at this conference are devoted to
different aspects of what we can call the ``\emph{standard}'' neutrino
oscillation scenario: only three neutrino flavors involved, no
interactions beyond those predicted by the Standard Model, and
neutrino conversion completely due to non-zero neutrino masses and
mixing.
In this talk I will instead focus on some of the \emph{alternative}
models which along the years have been proposed as possible
explanations of the various neutrino anomalies. Since none of these
models is presently able to account by itself for all the experimental
evidence, I will always consider the case where New Physics is
introduced \emph{in addition} to the conventional neutrino masses,
rather than \emph{in alternative} to them. In this context, I will
discuss the implications of each model for neutrino oscillations, the
bounds which can be put on its parameter space from the analysis of
present data, and the potentialities offered by futures experiments to
further improve these bounds.

The list of non-standard mechanisms for neutrino conversions proposed
so far is very large: it includes models of neutrino magnetic moment,
long-range leptonic forces, mass-varying neutrinos, violation of
fundamental principles, and much more. For definiteness and lack of
space, I will focus here only on three models, which in my view have
received most of the attention during the last few years: non-standard
interactions with matter, extra sterile neutrinos, and neutrino decay
and decoherence. Note that my approach in what follows will be purely
phenomenological: I will not make any reference to the theoretical
motivations of each model, focusing only on its experimental
implications.

\section{Non-standard interactions with matter}

The effective low-energy Lagrangian for neutrino interactions with
matter predicted by the Standard Model is:
\begin{equation} \label{eq:lagr-sm}
    \mathcal{L}_\text{SM}^\text{eff} = -2\sqrt{2} G_F
    \sum_{\beta} \LT(
    [ \bar\nu_\beta \gamma_\mu L \ell_\beta ]
    [ \bar{f} \gamma^\mu L f' ] + \text{h.c.} \RT)
    -2\sqrt{2} G_F \sum_{P,\beta} g_P^f
    [ \bar\nu_\beta \gamma_\mu L \nu_\beta ]
    [ \bar{f} \gamma^\mu P f ]
\end{equation}
where the first and the second term describe charged-current (CC) and
neutral-current (NC) interactions, respectively. Here $P \in \{ L,R
\}$, $(f,f')$ form an SU(2) doublet, and $g_P^f$ is the $Z$ coupling
to the fermion $f$.
Non-standard neutrino-matter interactions (NSI) can be introduced by
generalizing each term of Eq.~\eqref{eq:lagr-sm}. CC-like NSI are
severely constrained by their implications in the charged-lepton
sector, and although it has been shown that there is still room for
sizable effects at neutrino experiments~\cite{Huber:2002bi,
Kopp:2007mi, Kopp:2007ne, Ohlsson:2008gx}, they are usually ignored in
the literature, so I will not discuss them here. As for NC-like NSI, a
common parametrization is:
\begin{equation}
    \mathcal{L}_\text{NSI}^\text{eff} = - 2\sqrt{2} G_F
    \sum_{P,\alpha,\beta} \Eps_{\alpha\beta}^{fP} \,
    [ \bar\nu_\alpha \gamma_\mu L \nu_\beta ]
    [ \bar{f} \gamma^\mu P f ]
    \quad\text{with}\quad
    \Eps_{\beta\alpha}^{fP} = ( \Eps_{\alpha\beta}^{fP} )^*
\end{equation}
where $\Eps_{\alpha\beta}^{fP}$ denotes the strength of the
non-standard interactions between the neutrinos of flavors $\alpha$
and $\beta$ and the left-handed or right-handed components of the
fermion $f$.

\begin{table}\centering
    \caption{\label{tab:nsi}%
      Present bounds at 90\% CL on the NC-like NSI couplings
      $\Eps_{\alpha\beta}^{fP}$ from non-oscillation experiments.
      Limits have been obtained by varying each $\Eps_{\alpha\beta}$
      one at a time, with all the others set to zero.}
    \begin{tabular}{ccccc}
	\br
	Left-handed & Right-handed & Process & Experiment & Reference \\
	\br
	$-0.03 < \Eps_{ee}^{eL} < 0.08$ & $0.004 < \Eps_{ee}^{eR} < 0.15$
	& $\begin{aligned} \nu_e e &\to \nu e \\[-2mm]
	\bar{\nu}_e e &\to \bar{\nu} e \end{aligned}$
	& $\begin{aligned} &\text{LSND} \\[-2mm] &\text{Reactors} \end{aligned}$
	& \cite{Barranco:2005ps, Barranco:2007ej}
	\\[1mm]
	$-1 < \Eps_{ee}^{uL} < 0.3$ & $-0.4 < \Eps_{ee}^{uR} < 0.7$
	& $\nu_e q \to \nu q$ & CHARM
	& \cite{Davidson:2003ha}
	\\[1mm]
	$-0.3 < \Eps_{ee}^{dL} < 0.3$ & $-0.6 < \Eps_{ee}^{dR} < 0.5$
	& $\nu_e q \to \nu q$ & CHARM & \cite{Davidson:2003ha}
	\\
	\mr
	$|\Eps_{\mu\mu}^{eL}| < 0.03$ & $|\Eps_{\mu\mu}^{eR}| < 0.03$
	& $\nu_\mu e \to \nu e$ & CHARM II
	& \cite{Davidson:2003ha, Barranco:2007ej}
	\\[1mm]
	$|\Eps_{\mu\mu}^{uL}| < 0.003$ & $-0.008 < \Eps_{\mu\mu}^{uR} < 0.003$
	& $\nu_\mu q \to \nu q$ & NuTeV
	& \cite{Davidson:2003ha}
	\\[1mm]
	$|\Eps_{\mu\mu}^{dL}| < 0.003$ & $-0.008 < \Eps_{\mu\mu}^{dR} < 0.015$
	& $\nu_\mu q \to \nu q$ & NuTeV
	& \cite{Davidson:2003ha}
	\\
	\mr
	$-0.5 < \Eps_{\tau\tau}^{eL} < 0.2$ & $ -0.3 < \Eps_{\tau\tau}^{eR} < 0.4$
	& $e^+ e^- \to \nu \bar{\nu} \gamma$ & LEP
	& \cite{Berezhiani:2001rs, Barranco:2007ej}
	\\[1mm]
	$|\Eps_{\tau\tau}^{uL}| < 1.4$ & $|\Eps_{\tau\tau}^{uR}| < 3$
	& rad.~corrections & $\tau$ decay
	& \cite{Davidson:2003ha}
	\\[1mm]
	$|\Eps_{\tau\tau}^{dL}| < 1.1$ & $|\Eps_{\tau\tau}^{dR}| < 6$
	& rad.~corrections & $\tau$ decay
	& \cite{Davidson:2003ha}
	\\
	\br
	$|\Eps_{e\mu}^{eL}| < 0.0005$ & $|\Eps_{e\mu}^{eR}| < 0.0005$
	& rad.~corrections & $\mu\to 3e$
	& \cite{Davidson:2003ha}
	\\[1mm]
	$|\Eps_{e\mu}^{uL}| < 0.0008$ & $|\Eps_{e\mu}^{uR}| < 0.0008$
	& rad.~corrections & $\text{Ti}\, \mu \to \text{Ti}\, e$
	& \cite{Davidson:2003ha}
	\\[1mm]
	$|\Eps_{e\mu}^{dL}| < 0.0008$ & $|\Eps_{e\mu}^{dR}| < 0.0008$
	& rad.~corrections & $\text{Ti}\, \mu \to \text{Ti}\, e$
	& \cite{Davidson:2003ha}
	\\
	\mr
	$|\Eps_{e\tau}^{eL}| < 0.33$ & $|\Eps_{e\tau}^{eR}| < 0.28$
	& $\nu_e e \to \nu e$ & LEP+LSND+Rea
	& \cite{Berezhiani:2001rs, Barranco:2007ej}
	\\[1mm]
	$|\Eps_{e\tau}^{uL}| < 0.5$ & $|\Eps_{e\tau}^{uR}| < 0.5$
	& $\nu_e q \to \nu q$ & CHARM
	& \cite{Davidson:2003ha}
	\\[1mm]
	$|\Eps_{e\tau}^{dL}| < 0.5$ & $|\Eps_{e\tau}^{dR}| < 0.5$
	& $\nu_e q \to \nu q$ & CHARM
	& \cite{Davidson:2003ha}
	\\
	\mr
	$|\Eps_{\mu\tau}^{eL}| < 0.1$ & $|\Eps_{\mu\tau}^{eR}| < 0.1$
	& $\nu_\mu e \to \nu e$ & CHARM II
	& \cite{Davidson:2003ha, Barranco:2007ej}
	\\[1mm]
	$|\Eps_{\mu\tau}^{uL}| < 0.05$ & $|\Eps_{\mu\tau}^{uR}| < 0.05$
	& $\nu_\mu q \to \nu q$ & NuTeV
	& \cite{Davidson:2003ha}
	\\[1mm]
	$|\Eps_{\mu\tau}^{dL}| < 0.05$ & $|\Eps_{\mu\tau}^{dR}| < 0.05$
	& $\nu_\mu q \to \nu q$ & NuTeV
	& \cite{Davidson:2003ha}
	\\
	\br
    \end{tabular}
\end{table}    

In Table~\ref{tab:nsi} we summarize the present limits on
$\Eps_{\alpha\beta}^{fP}$ from various \emph{non-oscillation}
experiments. Clearly only the interactions of neutrinos with the
constituents of ordinary matter, $f = e, u, d$, are experimentally
accessible. As can be seen, the bounds are usually at the percent
level when a $\nu_\mu$ is involved, weak but still relevant (better
than unity) when a $\nu_e$ is present, and almost nonexistent for
$\nu_\tau$. Note that these limits have been obtained by varying each
$\Eps_{\alpha\beta}$ one at a time; in general, when correlations
among different $\Eps_{\alpha\beta}$ are included the bounds become
weaker~\cite{Barranco:2005ps, Barranco:2007ej}.

What can we learn on non-standard interactions from \emph{oscillation}
experiments? Neutrino production usually occurs through CC processes,
hence it is not affected by NC-like NSI. High-energy (above 100 MeV)
neutrinos are detected through the observation of the charged lepton
produced in CC interactions, but some solar neutrino experiment uses
signatures sensitive to NC processes (for example, $\nu + e \to \nu +
e$ elastic scattering in SK and Borexino, or $\nu + d \to \nu + p + n$
in SNO). As for neutrino propagation, in the presence of NSI an extra
term appears in the matter potential, $V_{\alpha\beta}^\text{NSI} =
\sqrt{2} G_F \sum_f N_f \Eps_{\alpha\beta}^{fV}$, with
$\Eps_{\alpha\beta}^{fV} = \Eps_{\alpha\beta}^{fL} +
\Eps_{\alpha\beta}^{fR}$. Hence neutrino oscillation experiments can
provide information on non-standard interactions. Note that due to the
very strong bounds on $\Eps_{\mu\mu}^{fP}$ and $\Eps_{e\mu}^{fP}$ (see
Table~\ref{tab:nsi}) it is common practice in numerical analyses to
assume $\Eps_{\mu\mu}^{fV} = 0$ and $\Eps_{e\mu}^{fV} = 0$ from the
very beginning.

The impact of NSI on solar neutrinos has been studied in detail. One
interesting fact is that the observed deficit of solar $\nu$ can be
perfectly explained by \emph{NSI only}, \Ie\ without the need of
mass-induced oscillations~\cite{Guzzo:2001mi, Gago:2001si}. However,
KamLAND is not affected by NSI and requires $\Dmq_{21} \ne 0$, hence
this pure NSI solution is no longer interesting. Combined
oscillation~+~NSI analyses of solar and KamLAND data have therefore
been performed~\cite{Miranda:2004nb, Guzzo:2004ue, Friedland:2004pp},
but the bounds they impose on the NSI parameters are very weak. Hence
at present no interesting information on NSI can be extracted from
solar data. Note, however, that none of these analyses takes into
account the Borexino result presented at this conference. 

The situation for atmospheric neutrinos is quite different. Although a
complete three-neutrino analysis cannot be done due to the very high
number of parameters involved, partial analyses have been performed.
NSI in the $\mu - \tau$ sector ($\Eps_{e\alpha} = \Eps_{\alpha e} =
0$) have been studied in~\cite{Fornengo:2001pm,
GonzalezGarcia:2004wg}, finding that the bounds on the NSI parameters
implied by atmospheric data are very strong. The most recent fit
including also accelerator experiments~\cite{GonzalezGarcia:2007ib}
gives $|\Eps_{\mu\tau}^V| \le 0.038$ and $|\Eps_{\tau\tau}^V| \le
0.12$ at 90\% CL, with $\Eps_{\alpha\beta}^V \equiv
\Eps_{\alpha\beta}^{eV} + 3 \Eps_{\alpha\beta}^{uV} + 3
\Eps_{\alpha\beta}^{dV}$ (the factor 3 is the approximate $N_u/N_e$
and $N_d/N_e$ ratio in the Earth matter). Note that the bounds on
$\Eps_{\tau\tau}$ listed in Table~\ref{tab:nsi} are more than one
order of magnitude weaker. NSI in the $e - \tau$ sector
($\Eps_{\mu\alpha} = \Eps_{\alpha\mu} = 0$) have been considered
in~\cite{Friedland:2004ah, Friedland:2005vy, Friedland:2006pi}, and in
this case the sensitivity to the NSI parameters is much poorer. In
particular, the bound on $\Eps_{e\tau}^V$ is of order unity, hence
worse than those imposed by non-oscillation experiments. As for the
bound on $\Eps_{\tau\tau}^V$ previously quoted, it still hold provided
that it is reinterpreted as a bound on the combination
$\Eps_{\tau\tau}^V - |\Eps_{e\tau}^V|^2 / (1 + \Eps_{ee}^V)$. This
demonstrates that correlations among different parameters can have
very important consequences.

Let us now turn to \emph{future} experiments. The potentialities of
neutrino factories for the determination of NSI parameters was first
considered in~\cite{Huber:2001zw}, where it was shown that they will
provide complementary information to atmospheric neutrino experiments.
However, it was soon realized that due to degeneracies between NSI and
oscillation parameters the sensitivity of a neutrino factory to
$\theta_{13}$ could be seriously spoiled in the presence of
NSI~\cite{Huber:2001de, Huber:2002bi}. The situation became less
dramatic if data from two different baselines were
combined~\cite{Huber:2001de}. More recent studies confirm these
results, and show that while an experiment with a single baseline is
strongly affected by degeneracies~\cite{Kopp:2007mi}, a two-baseline
configuration (\textit{e.g.}, 3000--4000 and 7000--7500 km) can
simultaneously provide a robust determination of the oscillation
parameters and strong constraints on non-standard
interactions~\cite{Ribeiro:2007ud, Kopp:2008ds, Winter:2008eg}.

The potentialities of forthcoming and long-term facilities have been
discussed in a number of papers.
Coherent scattering of low energy neutrinos is very sensitive to NSI
with quarks, and offers the possibility to improve dramatically the
bounds on $\smash{\Eps_{ee}^{qV}}$ and
$\smash{\Eps_{e\tau}^{qV}}$~\cite{Barranco:2005yy, Barranco:2007tz}.
In the context of solar neutrinos, the precise measurement of the
$^7$Be line in Borexino can provide very important information on
NSI~\cite{Berezhiani:2001rt}.
The sensitivity to $\theta_{13}$ of MINOS~\cite{Blennow:2007pu} and of
beta-beams~\cite{Adhikari:2006uj} can be seriously spoiled if NSI are
present; this problem, which affects all single-baseline experiments,
can be efficiently resolved by the combination with a reactor
experiment~\cite{Kopp:2007ne}. OPERA is too small to provide any
useful information on $\Eps_{e\tau}^V$ and
$\Eps_{\tau\tau}^V$~\cite{EstebanPretel:2008qi}, but it may help in
the determination of $\Eps_{\mu\tau}^V$~\cite{Blennow:2008ym}, to
which T2KK will also have a good sensitivity~\cite{Ribeiro:2007jq}.

\section{Models with extra sterile neutrinos}

In April 2007 the MiniBooNE collaboration released their first
data~\cite{AguilarArevalo:2007it} on a search for $\nu_\mu \to \nu_e$
appearance with a baseline of 540~m and a mean neutrino energy of
about 700~MeV. This experiment did not find any signal compatible with
two-neutrino oscillations, however an unexplained $3.6\sigma$ excess
was observed in the low-energy region.
The primary purpose of this experiment was to test the evidence of
$\bar\nu_\mu \to \bar\nu_e$ transitions reported by the LSND
experiment at Los Alamos~\cite{Aguilar:2001ty} with a very similar
$L/E$ range. Since the mass-squared differences required to explain
the solar, atmospheric and LSND experimental results in terms of
neutrino oscillations differ from one another by various orders of
magnitude, there is no consistent way to reconcile these three signals
using only oscillations among the three known neutrinos. A popular way
to solve the LSND problem is to invoke an extension of the
three-neutrino mixing scenario, where at least three mass-square
differences are available due to the introduction of one or more extra
(sterile) neutrino states. An updated analysis of such models
including also the MiniBooNE result was presented in
Ref.~\cite{Maltoni:2007zf}. It was found that:
\begin{itemize}
  \item four-neutrino models are ruled out since (a) the don't allow to
    account for the low energy event excess in MiniBooNE, (b)
    MiniBooNE result cannot be reconciled with LSND, and (c) there is
    severe tension between \emph{appearance} ($\nu_e \to \nu_\mu$ and
    $\nu_\mu \to \nu_e$) and \emph{disappearance} ($\nu_e \to \nu_e$
    and $\nu_\mu \to \nu_\mu$) experiments;
    
  \item five-neutrino models provide a nice way out for problems (a)
    and (b), but fail to resolve (c);
    
  \item six-neutrino models do not offer qualitatively new
    effects with respect to the previous case.
\end{itemize}
In all the cases the authors find severe tension between different
sub-samples of the data, hence they conclude that at the light of
present experimental results it is \emph{not} possible to explain the
LSND evidence in terms of sterile neutrinos.

Since the existence of sterile neutrinos beyond the three known ones
is a very interesting issue by itself, it is worth to consider it
irrespectively of whether the LSND anomaly is confirmed or not. A
number of studies discussing the sensitivity of future experiments to
extra sterile states have been presented, in the context of
Opera~\cite{Donini:2007yf}, of neutrino factories~\cite{Barger:2000ax,
Donini:2001xp, Dighe:2007uf}, of $\beta$-decay
experiments~\cite{Goswami:2007kv}, and of neutrino
telescopes~\cite{Awasthi:2007az, Choubey:2007ji, Donini:2008xn}. It
should be noted that all these works still assume that the extra
neutrinos are heavier than about 1~eV, whereas once LSND is dropped
there is no reason to make any assumption on the mass of the sterile
states. However, this general case has been considered only in a very
few works~\cite{deHolanda:2003tx, Barger:2004db, Cirelli:2004cz,
deGouvea:2008qk}.

\section{Neutrino decay and decoherence}

Although the theoretical motivations for neutrino decay and neutrino
decoherence are very different, they are characterized by the same
phenomenological signature: an exponential damping of the flavor
conversion probabilities. Hence we will discuss them together.

Concerning neutrino decay, from the phenomenological point of view we
should distinguish two possible situation: $\nu_i \to
\stackrel{\smash{\scriptscriptstyle(-)}}{\nu}_{\hspace{-3pt}j} + X$,
\Ie\ when the decay product include one (or more) detectable
neutrinos, and $\nu_i \to X$, \Ie\ when the decay products are
completely invisible. In the first case, the energy distribution of
the daughter neutrino(s) is model-dependent, whereas in the second
case the process is completely described by the \emph{neutrino
lifetime} $\tau_i$ and the evolution equation is obtained by adding an
imaginary part to the vacuum Hamiltonian, $H_0^m \to H_0^m - i
\Gamma_0^m$, with
\begin{equation}
    H_0^m = \frac{1}{2 E_\nu} \diag
    \LT( 0, \Dmq_{21}, \Dmq_{31} \RT)
    \quad\text{and}\quad
    \Gamma_0^m = \frac{1}{2 E_\nu} \diag
    \LT( \frac{m_1}{\tau_1},
    \frac{m_2}{\tau_2}, \frac{m_3}{\tau_3} \RT).
\end{equation}
Note that since the neutrino masses $m_i$ are unknown, one typically
quotes $\tau_i/m_i$ as the neutrino lifetime. Interference effects
between oscillations and decay~\cite{Lindner:2001fx} are usually
neglected.

In general, the strength of the bounds on the neutrino lifetimes
increases with the baseline of the experiment imposing them. The best
limit follows from the observation of neutrino events associated with
the explosion of SN1987A, which leads to a bound $\tau_1/m_1 \gtrsim
10^5~\text{s/eV}$~\cite{Hirata:1987hu} on the lifetime of the lightest
neutrino state $\nu_1$. 
Bounds on $\nu_2$ lifetime are much weaker, and are dominated by solar
neutrino data. For the case of invisible decay, the non-observation of
$\nu_2$ disappearance implies $\tau_2/m_2 \gtrsim 8.7 \times
10^{-5}~\text{s/eV}$ at 99\% CL~\cite{Joshipura:2002fb,
Bandyopadhyay:2002qg}, although it has been pointed out that this
limit may not hold for quasi-degenerate
neutrinos~\cite{Beacom:2002cb}. As for decay modes with secondary
$\bar{\nu}_e$ appearance, KamLAND~\cite{Eguchi:2003gg} and
SNO~\cite{Aharmim:2004uf} performed dedicated searches for
antineutrinos coming from the Sun, yielding $\tau_2/m_2 > 1.1 \times
10^{-3}~\text{s/eV}$ for hierarchical masses and $\tau_2/m_2 > 6.7
\times 10^{-2}~\text{s/eV}$ for quasi-degenerate
masses~\cite{Eguchi:2003gg}.
Limits on $\nu_3$ lifetime follow from the analysis of atmospheric and
long-baseline neutrino data, and given the much shorter path length
they are considerably weaker than those quoted so far. A pure decay
solution ($\Dmq_{31} = 0$) of the atmospheric deficit was still
possible until a few years ago~\cite{Barger:1999bg}, but it is now
ruled out by Super-Kamiokande~\cite{Ashie:2004mr}. Interestingly,
atmospheric data also admit an hybrid oscillation~+~decay
solution~\cite{Choubey:1999ir} with $\tau_3/m_3 \simeq 2.6 \times
10^{-12}~\text{s/eV}$ and $\theta_{23} = 34^\circ$, which is however
ruled out by MINOS, leading to the bound $\tau_3/m_3 > 2.9\times
10^{-10}~\text{s/eV}$ at 90\%~CL~\cite{GonzalezGarcia:2008ru}.

Neutrino decoherence can arise from a number of very different
phenomena: averaging due to finite detector resolution, finite-size of
the neutrino wave-packet, quantum-gravity interactions of neutrinos
with the space-time ``foam'', and so on. Phenomenologically,
decoherence lead to the appearance of a damping term
$\mathcal{D}[\rho]$ in the evolution equation of the neutrino density
matrix $\rho$: $d\rho / dt = -i [H, \rho] - \mathcal{D}[\rho]$. The
specific form of $\mathcal{D}[\rho]$ is model-dependent, however it is
common in the literature to make a number of conservative assumptions
(complete positivity, unitarity, increase of the Von Neumann entropy,
and conservation of energy in vacuum) which lead to the simple
expression $\mathcal{D}[\rho] = \sum_\ell [D_\ell, [D_\ell, \rho]]$
with $D_\ell = \diag(d_{\ell 1}, d_{\ell 2}, d_{\ell 3})$ in the
vacuum mass basis. In this case the evolution equation in vacuum can
be solved analytically, and three new parameters $\gamma_{ji} =
\sum_\ell (d_{\ell j} - d_{\ell i})^2$ appear in addition to the usual
ones. Note that in general $\gamma_{ji}$ can depend on the neutrino
energy.

\begin{table}\centering
    \caption{\label{tab:decoh}%
      Present bounds on neutrino decoherence in different two-neutrino
      oscillation channels, assuming a power law dependence
      $\gamma(E_\nu) = \kappa_n (E_\nu / \text{GeV})^n$. Bounds marked
      as ``old'' are in the same units as the corresponding ``new''
      ones, and are taken from~\cite{Lisi:2000zt, Fogli:2003th}.}
    \begin{tabular}{c@{\hspace{10mm}}c@{~}l}
	\br
	$\nu_e \to \nu_x$ (95\% CL) &
	\multicolumn{2}{c}{$\nu_\mu \to \nu_\tau$ (90\% CL)} \\
	\br
	$\kappa_{-2}^\text{sol} < 8.1\times 10^{-29}$~GeV &
	$\kappa_{-2}^\text{atm} < 1.9\times 10^{-22}$~GeV
	\\[1mm]
	$\kappa_{-1}^\text{sol} < 7.8\times 10^{-27}$~GeV &
	$\kappa_{-1}^\text{atm} < 1.2\times 10^{-22}$~GeV &
	(old: 20)
	\\[1mm]
	$\kappa_{0\hphantom{+}}^\text{sol} < 6.7\times 10^{-25}$~GeV &
	$\kappa_{0\hphantom{+}}^\text{atm} < 2.7\times 10^{-24}$~GeV &
	(old: 35)
	\\[1mm]
	$\kappa_{+1}^\text{sol} < 5.8\times 10^{-23}$~GeV &
	$\kappa_{+1}^\text{atm} < 3.8\times 10^{-27}$~GeV
	\\[1mm]
	$\kappa_{+2}^\text{sol} < 4.7\times 10^{-21}$~GeV &
	$\kappa_{+2}^\text{atm} < 2.4\times 10^{-30}$~GeV &
	(old: 900)
	\\
	\br
    \end{tabular}
\end{table}

Phenomenological analyses performed so far focus on two-neutrino
oscillations, for which only one $\gamma$ at a time is relevant.
Decoherence involving $\nu_e$ is constrained by
KamLAND~\cite{Schwetz:2003se} as well as solar neutrino
data~\cite{Fogli:2007tx}. For KamLAND, the relevant probability can be
written explicitly,
\begin{equation}
    P_{ee} = 1 - \frac{1}{2} \sin^2(2\theta) \LT[ 1 -
    e^{-\gamma_\text{sol} L} \cos\LT(\frac{\Dmq L}{2 E_\nu} \RT) \RT],
\end{equation}
whereas for solar neutrinos matter effects cannot be neglected. The
limits implied by a combined analysis of both
experiments~\cite{Fogli:2007tx} assuming a power law dependence
$\gamma_\text{sol}(E_\nu) = \kappa_n^\text{sol} (E_\nu /
\text{GeV})^n$ are listed in Table~\ref{tab:decoh}.
Decoherence in the $\nu_\mu \to \nu_\tau$ channel has been studied in
the context of atmospheric and accelerator neutrino experiments.
Similarly to the case of neutrino decay, a pure decoherence solution
was originally allowed~\cite{Lisi:2000zt}, but it is now ruled out at
more than $3\sigma$~\cite{Ashie:2004mr}. A combined
oscillation~+~decoherence fit for $\gamma_\text{atm}(E_\nu) =
\kappa_n^\text{atm} (E_\nu / \text{GeV})^n$ was first presented
in~\cite{Lisi:2000zt, Fogli:2003th}; updated results including the
latest SK-I and SK-II data as well as K2K and MINOS are reported in
Table~\ref{tab:decoh}.

Various attempts have been made to explain the LSND results in terms
of neutrino decay or decoherence in combination with oscillations, but
usually other kinds of New Physics are needed as well: for example,
decay~+~sterile neutrinos~\cite{PalomaresRuiz:2005vf},
decoherence~+~CPT-violation~\cite{Barenboim:2004wu}, decoherence with
unusual $L$ dependence~\cite{Barenboim:2006xt}, and so on. A very
interesting model recently proposed~\cite{Farzan:2008zv} involves
oscillations plus decoherence in the general three-neutrino scenario:
assuming $\gamma_{21} = 0$ and $\gamma_{31}(E_\nu) =
\gamma_{32}(E_\nu) = \kappa_{-4}^\text{atm} (E_\nu /
\text{GeV})^{-4}$, this model succeeds to reconcile all the
experimental evidence, except for the MiniBooNE low-energy excess,
provided that $\kappa_{-4}^\text{atm} = 1.7\times 10^{-23}$~GeV and
$\sin^2\theta_{13} > (2.6\pm 0.8) \times 10^{-3}$.

Only a few studies have been performed to investigate the sensitivity
of future neutrino facilities to neutrino decay and decoherence.
In~\cite{Blennow:2005yk} it was shown that decoherence effects can
fake the determination of $\theta_{13}$ at reactor experiments, and
that a neutrino factory can easily identify the presence of neutrino
decay, whereas its ability to recognize decoherence depends on the
specific shape of $\gamma(E_\nu)$.
In~\cite{Mavromatos:2007hv} it was found that the bounds on
decoherence parameters which can be put by CNGS and T2K are comparable
with those derived from atmospheric neutrinos. A similar result also
holds for T2KK~\cite{Ribeiro:2007jq}, which in the context of
decoherence models it is shown to be systematically better than the
separate Kamioka-only and Korea-only configurations.
On the other hand, the potentialities of future neutrino telescopes to
detect decay and decoherence signatures have received considerable
attention. 
Concerning neutrino decay~\cite{Beacom:2002vi, Beacom:2003zg,
Meloni:2006gv, Maltoni:2008jr}, due to the extremely long distance
traveled by astrophysical neutrinos their sensitivity to the neutrino
lifetime is many orders of magnitude larger than conventional
ground-based experiments. Moreover, neutrino decay can break the
$1:1:1$ flavor ratio expected from a $\pi$-decay source, hence opening
the possibility to measure oscillation parameters at neutrino
telescopes~\cite{Maltoni:2008jr}.
As for decoherence, under our restrictive assumptions it is
indistinguishable from averaged oscillations, however more general
scenarios predicting unique signatures have been
considered~\cite{Hooper:2004xr, Anchordoqui:2005gj}.

\section{Conclusions}

In this talk I have discussed the phenomenological implications of
different non-standard mechanisms for neutrino conversion. I have
focused on three specific cases: non-standard neutrino interactions
with matters, models with extra sterile neutrinos, and neutrino decay
and decoherence. For what concerns non-standard interactions, we have
shown that present bounds on NSI parameters are affected by strong
degeneracies, which could spoil the sensitivity to $\theta_{13}$ of
future long-baseline experiment and neutrino factories, but which can
be efficiently resolved by the combination of experiments with two
different baselines. Concerning sterile neutrino models, we have
proved that none of them succeed in reconciling LSND with the results
of the other neutrino oscillation experiments. As for neutrino decay
and decoherence, we have reviewed and updated the present limits on
the damping parameters, pointing out that future reactor and
accelerator facilities can further enhance these limits and that
neutrino decay can have non-trivial implications for neutrino
telescopes.

\section*{Acknowledgments}

Work supported by MICINN through the Ram\'on y Cajal program and through
the national project FPA2006-01105, and by the Comunidad Aut\'onoma de
Madrid through the HEPHACOS project P-ESP-00346.

\section*{References}

\providecommand{\newblock}{}

\end{document}